\documentclass{article}
\usepackage{graphicx}
\usepackage[english]{babel}

\title{
\includegraphics[width=0.35\textwidth]{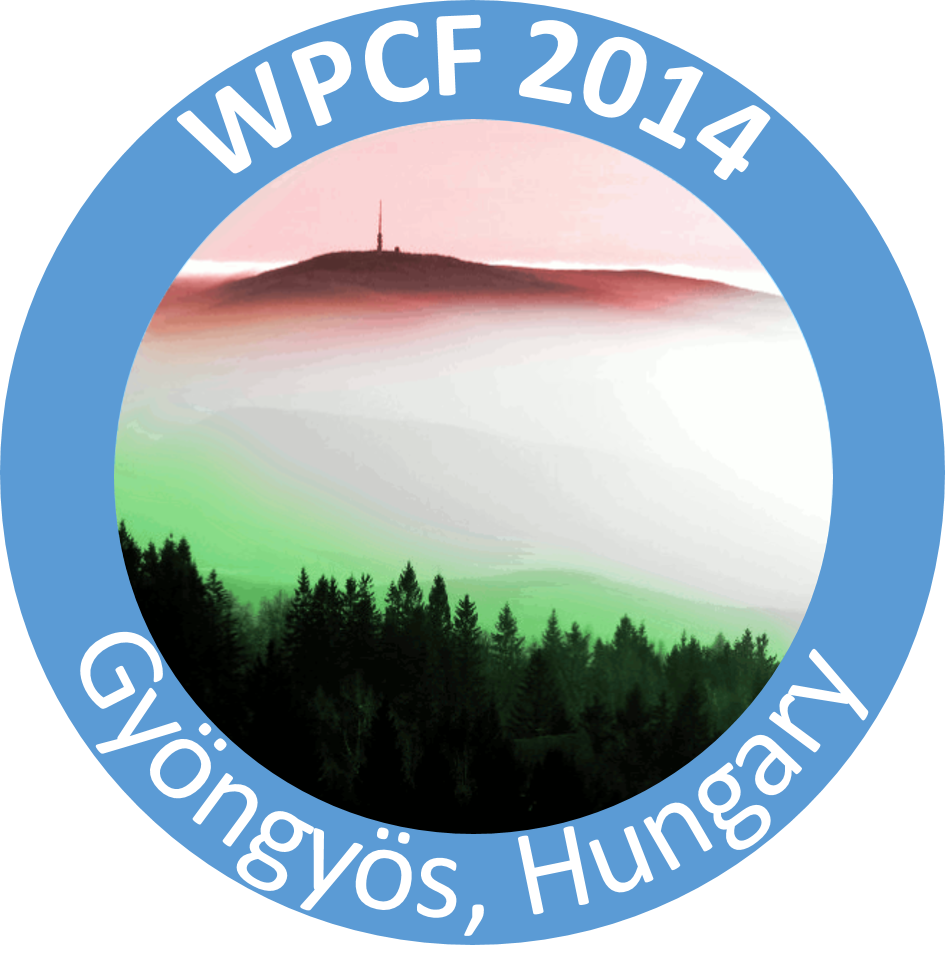}\\[1cm]
Charge splitting of directed flow
and 
charge-dependent effects
 in
pion spectra
in heavy ion collisions}
\author{{A. Rybicki$^1$, A. Szczurek$^{1,2}$, M. K\l{}usek-Gawenda$^1$, M. Kie\l{}bowicz$^3$}\\[1ex]
$^1$H.Niewodnicza\'nski
Institute of Nuclear Physics, Polish Academy\\ of Sciences, Radzikowskiego 152, 31-342 Krak\'ow, Poland\\
$^2$University of Rzesz\'ow, Rejtana 16, 35-959 Rzesz\'ow, Poland\\
$^3$The Tadeusz Ko\'sciuszko Cracow University of Technology,\\
Warszawska 24, 31-155 Krak\'ow, Poland\\
}

\begin{document}

\fontfamily{lmss}\selectfont
\maketitle

\begin{abstract}
The large and rapidly varying electric and magnetic fields induced by the 
spectator systems moving at ultrarelativistic velocities induce a charge splitting of directed flow, $v_1$, of positive and negative pions in the final state of the heavy ion collision. The same effect results in a very sizeable distortion of charged pion spectra as well as
ratios of charged pions ($\pi^+/\pi^-$) emitted at high values of rapidity. Both phenomena are sensitive to the actual distance between the pion emission site and the spectator system. This distance $d_E$ appears to decrease with increasing rapidity of the pion, and comes below $\sim$1~fm for pions emitted close to beam rapidity. In this paper we discuss how these findings can shed new light on 
the space-time evolution of pion production as a function of rapidity, and on the longitudinal evolution of the system created in heavy ion collisions.
\end{abstract}

\section{Introduction}

The presence of large and rapidly varying electric and magnetic fields in relativistic heavy ion collisions results in charge-dependent effects, visible in a series of observables in the final state of the collision. These effects can be used as a new source of information on the space-time evolution of the non-perturbative process of particle production, and on the space-time properties of the system created in the heavy ion collision.
To give one example, in 2007 we demonstrated that the distortion which the electromagnetic repulsion (attraction) of positive (negative) pions induced on charged pion ($\pi^+/\pi^-$) ratios brought new information on the space-time scenario of fast pion production~\cite{Rybicki-coulomb}. In recent years, the general problematics of electromagnetically-induced effects in ultrarelativistic heavy ion reactions was subject of an important theoretical and experimental interest~\cite{kharzeev2014,voloshin2014,hirono2012,workshop} as it was connected to very interesting phenomena like the chiral magnetic effect~(CME~\cite{cme,cme1}).

In the present paper we review our earlier studies of the electromagnetic distortion of charged pion spectra in the context of our more recent findings on the influence of spectator-induced $\vec{E}$ and $\vec{B}$ fields on the azimuthal anisotropies of charged pions. Special attention is put on tracing the utility of both observables for studying the longitudinal evolution of the expanding matter created in the collision. 
A phenomenological model analysis is presented, aimed at explaining the space-time features of pion production which we deduced from the observed electromagnetic phenomena. 

\section{Charged-dependent effects in pion spectra at the SPS}

\begin{figure}[h]
\begin{center}
\includegraphics[width=0.8\textwidth]{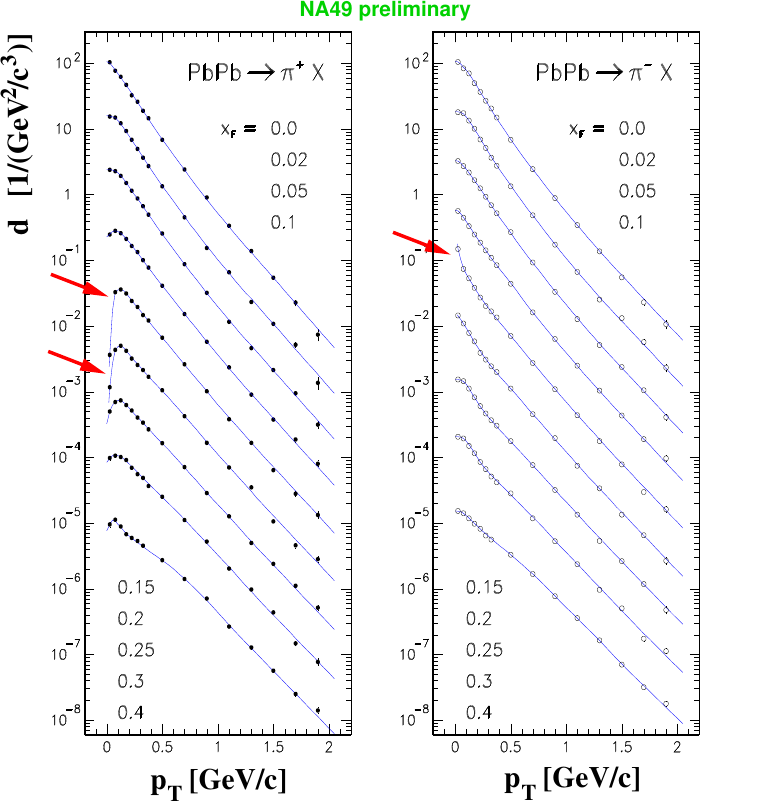}\\[-0.0cm]
\caption{Distibutions of invariant density $d=E\frac{d^3n}{dp^3}$ of positively and negatively charged pions produced in peripheral Pb+Pb collisions at $\sqrt{s_{NN}}=17.3$~GeV. The pion invariant density is drawn as a function of transverse momentum in fixed bins of $x_F$ as marked from top to bottom. The subsequent distributions are consecutively multiplied by 0.2. The arrows point at the regions where the distortion induced by the spectator EM-field is most visible. From~\cite{Rybicki-epshep}.}
\label{fig1a}
\end{center}
\end{figure}

The relatively moderate collision energy range available to the SPS makes corresponding fixed-target experiments suitable for studying the electromagnetic influence of the spectator system on charged particle spectra in a large range of available rapidity. Importantly, this includes the region of very low transverse momenta where the corresponding effects are expected to be largest. A detailed double-differential study of $\pi^+$ and $\pi^-$ densities as a function of longitudinal and transverse pion momentum is presented in Fig.~\ref{fig1a}. The NA49 experimental data cover, in the longitudinal direction expressed in terms of the c.m.s. Feynman variable $x_F=2p_L/\sqrt{s_{NN}}$, the whole region from ``mid-rapidity'' ($x_F=y=0$) up to $x_F=0.4$ which is about one unit above beam rapidity at lowest transverse momenta. The smooth exponential-like shape of the transverse momentum distribution gets visibly distorted in the region of low $p_T$, where a dramatic decrease of invariant $\pi^+$ density and an accumulation of $\pi^-$ density is apparent as indicated by the arrows. This ``deformation'' is caused by the spectator system, which modifies the trajectories of charged pions by means of its space- and time-dependent $\vec{E}$ and $\vec{B}$ fields. 

The ratio of $\pi^+$ over $\pi^-$ density, Fig.~\ref{fig1}(a), appears particularly sensitive to the spectator-induced electromagnetic field in the region of higher rapidity ($x_F>0.1$) and lower transverse momenta. Here, a deep two-dimensional ``valley'' is apparent with the $\pi^+/\pi^-$ ratio approaching zero in the region $y\approx y_\mathrm{beam}$ ($x_F=0.15=m_\pi/m_N$ at low $p_T$). Note that with the Pb nucleus composed of 39\% protons over 61\% neutrons, this implies breaking of isospin symmetry which unequivocally confirms the electromagnetic origin of the observed effect. 
Quantitatively, this is confirmed in Fig.~\ref{fig1}(b), where the observed distortion can be fairly well described by means of a simple two-spectator model with the two spectators assumed as Lorentz-contracted homegenously charged spheres, and isospin effects being taken into account~\cite{ismd2012}. It is important to underline that the unique free parameter in the model is the distance $d_E$, in the longitudinal direction, between the pion emission point and the center of the spectator system. The reasonable agreement between data and model demonstrated in Figs~\ref{fig1}(a),(b) is obtained for values of $d_E$ in the range of 0.5~-~1~fm~\cite{ismd2012}; different values of $d_E$ lead to different detailed shapes of the distortion of $\pi^+/\pi^-$ ratios as described in~\cite{Rybicki-coulomb}.

\section{Directed flow}

\begin{figure}[t]
\begin{center}
\includegraphics[width=0.9\textwidth,height=0.5\textheight]{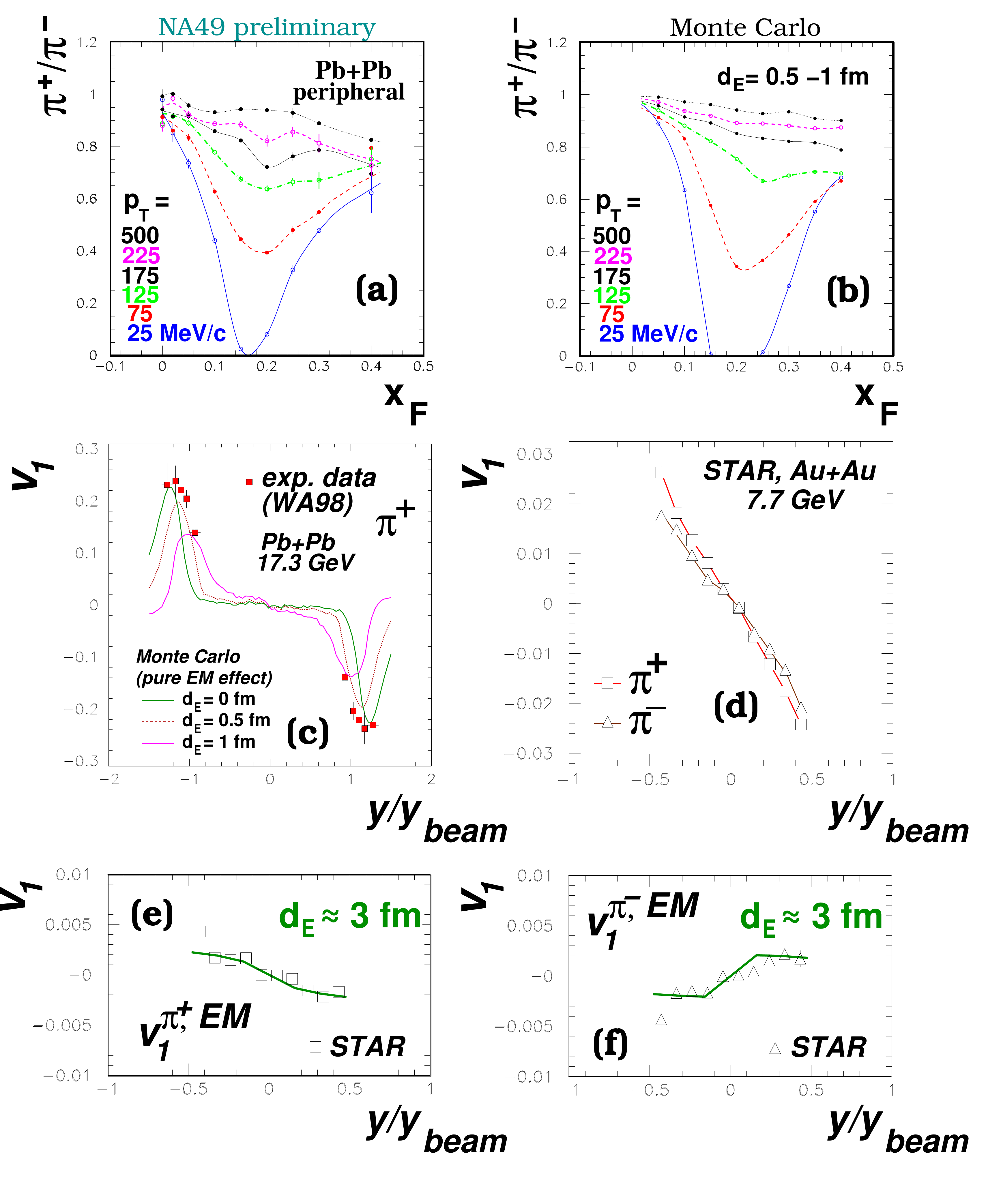}\\[-0.4cm]
\caption{(a) Ratio of charged pions emitted in peripheral Pb+Pb collisions at $\sqrt{s_{NN}}=17.3$~GeV, (b) model simulation of this ratio as described in the text, (c)~our Monte Carlo prediction for the (pure) electromagnetically-induced directed flow of positive pions, compared to the data from the WA98 experiment~\cite{wa98}, (d) directed flow of charged pions in intermediate centrality Au+Au collisions~\cite{star2014}, (e), (f) electromagnetic component of $\pi^+$ and $\pi^-$ directed flow, extracted from STAR data~\cite{star2014} and compared to our simulation made assuming $d_E\approx 3$~fm. From:~\cite{Rybicki-epshep}~(panels a,b),~\cite{Rybicki-v1} (panel c),~\cite{Rybicki-meson2014} (panels d,e,f).}
\label{fig1}
\end{center}
\end{figure}

In full analogy to charged pion ratios, the {\em directed flow} of charged pions emitted close to beam rapidity is also strongly affected by spectator-induced EM effects. This is shown in Fig.~\ref{fig1}(c) where our prediction for a {\em purely electromagnetic effect} on the directed flow $v_1$ of positive pions 
is shown for three different values of the distance $d_E$: 0, 0.5 and 1 fm. As it can be seen in the figure, our Monte Carlo calculation shows that very large values of directed flow can be induced by the sole effect of electromagnetic repulsion of positive pions by the spectator system.
Our prediction
is compared to the measurements provided by the WA98 Collaboration at the same energy, $\sqrt{s_{NN}}=17.3$~GeV~\cite{wa98}. This comparison indicates that a very sizeable part of positive pion directed flow in the region close to beam/target rapidity can in fact come from the electromagnetic origin. At the same time, the WA98 experimental data apparently constrain the possible values of the distance $d_E$, yielding the possible range of $d_E$ from 0 up to 1~fm. Thus consistently from both observables ($\pi^+/\pi^-$ ratios, Fig.~\ref{fig1}(a) and directed flow, Fig.~\ref{fig1}(c)), the longitudinal distance between the actual pion emission site and the center of the spectator system appears quite small, in the range below 1~fm. This small distance is to be viewed with respect to the longitudinal extent of the Lorentz-contrated spectator system which is itself of the order of about 1~fm at this collision energy.

The situation changes significantly when passing to pions produced close to {\em central} rather than {\em beam} rapidity. Here experimental data on intermediate centrality Au+Au reactions exist from the STAR experiment at RHIC~\cite{star2014} at different collision energies (from $\sqrt{s_{NN}}=7.7$ up to $200$~GeV). The directed flow of positive and negative pions at the lowest available energy is presented in Fig.~\ref{fig1}(d). A {\em charge splitting} is apparent between $\pi^+$ and $\pi^-$. As shown in Figs~\ref{fig1}(e),(f), the latter splitting can again be understood as a spectator-induced EM effect, provided that a value of $d_E$ far larger than in the preceding case, $d_E\approx 3$~fm, is assumed.

\section{Space-time picture of the collision}

\begin{figure}[h!]
\begin{center}
\includegraphics[width=0.8\textwidth,height=0.55\textheight]{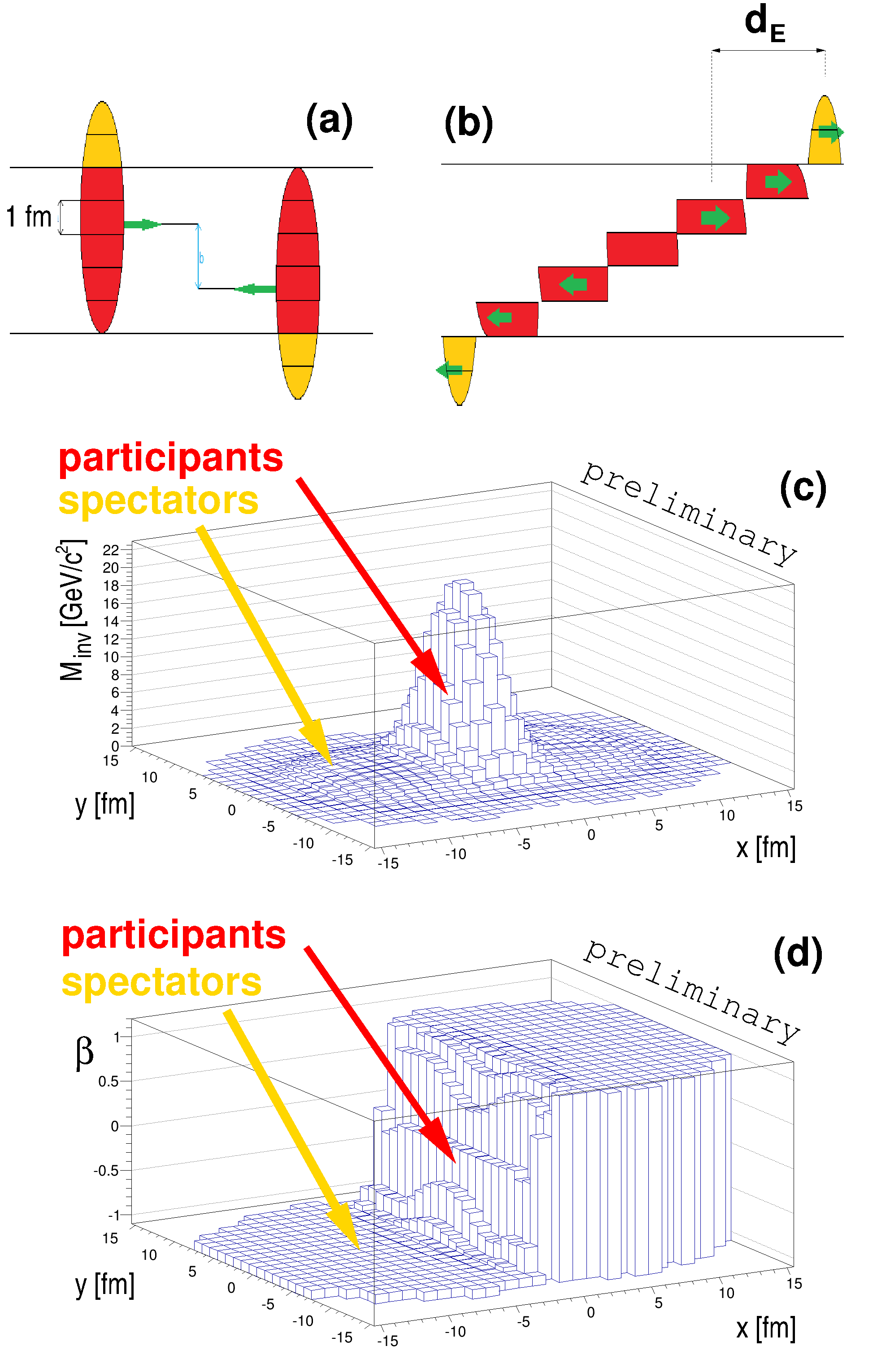}\\[-0.4cm]
\caption{Model study of the kinematical characteristics of matter created in the peripheral Pb+Pb collision at $\sqrt{s_{NN}}=17.3$~GeV. (a)~Subdivision of the nuclear matter distribution into longitudinal ``strips''. (b)~Kinematical characteristics of the ``strips'' as a function of their position in the perpendicular plane; the distance $d_E$ is indicated in the plot. (c)~Invariant mass of the ``strips''  projected in the perpendicular $(x,y)$ plane, where $x$ is the direction of the impact parameter vector. (d)~Longitudinal velocity $\beta$ of the ``strips'' as a function of their position. The ``hot'' participant and ``cold'' spectator regions are indicated in the plots.}
\label{fig2}
\end{center}
\end{figure}

This apparent sensitivity of the electromagnetic distortion of final state charged pion ratios and directed flow to the distance between the pion formation zone and the spectator system provides, in the opinion of the authors, a completely new and very welcome tool for studying the space-time evolution of charged particle production in the soft sector of ultrarelativistic heavy ion collisions. Specifically, the elongation of the distance $d_E$ with decreasing pion rapidity is the reflection of the longitudinal evolution of the system created in the collision.
% as depicted in Fig~\ref{fig1}(g). 
%Summing up the findings from the precedent section we have:
Summing up the findings from the precedent section, in our studies we obtained:
\begin{itemize}
\item[$\bullet$] $d_E\leq 1$~fm for pions moving at rapidities comparable to $y_\mathrm{beam}$ (from our 
study
%analysis 
based on 
%of
NA49~\cite{Rybicki-epshep} and WA98~\cite{wa98} data);
\item[$\bullet$] $d_E\approx 3$~fm for pions moving at central rapidities ($-1<y<1$, from our 
%analysis 
study
based on
%of 
STAR data~\cite{star2014}).
\end{itemize}
While the mere fact that $d_E$ evolves with pion rapidity is simply the confirmation of the expansion of the system in the longitudinal direction, the latter is, especially at high pion rapidities, poorly known to hydrodynamical calculations due to the presence of a sizeable baryochemical potential~\cite{Ryblewski}, and difficult to access experimentally e.g. in LHC experiments (in contrast to SPS energies where the NA49 and NA61/SHINE experiments cover the whole region from $y=0$ to $y=y_\mathrm{beam}$ and above in the collision c.m.s.~\cite{na61-2014}). 

In the present section we discuss this issue in the context of energy-momentum conservation in the initial state of the collision, in a model proposed by A.S. The spatial nuclear matter distribution in the volume of the two colliding nuclei is considered in a two-dimensional $(x,y)$ projection perpendicular to the collision axis; peripheral Pb+Pb collisions at top SPS energy are presented in Fig.~\ref{fig2}(a). The resulting ``strips'' of highly excited nuclear (or partonic) matter, Fig.~\ref{fig2}(b), define the kinematical properties of the longitudinal expansion of the system as a function of collision geometry. These are shown in Figs~\ref{fig2}(c) and (d) in the perpendicular $(x,y)$ plane. For the peripheral collision considered here, the overall energy available for particle production (invariant mass of the ``strips'' as defined assuming local energy-momentum conservation) has a well-defined ``hot'' peak at mid-distance between the centers of the two nuclei, and gradually decreases when approaching each of the two ``cold'' spectator systems. On the other hand, the longitudinal velocity $\beta$ of the ``strips'' depends strongly on their position in the $(x,y)$ plane. A careful comparison of Figs~\ref{fig2}(c)-(d) shows that significantly excited volume elements of the longitudinally expanding system can move at very large longitudinal velocities, comparable to that of the spectator system. Assuming a given proper hadronization time of the different volume elements, a natural picture emerges. Pions produced at high rapidity (dominantly from ``strips'' moving at large 
%longitudinal velocities) 
values of the longitudinal velocity $\beta$)
will emerge at a small distance from the ``cold'' spectator systems; these originating from ``hot'' central ``strips'', at smaller values of $y$, will evidently show up at larger values of the distance $d_E$.

\section{Conclusions}

Altogether, we conclude that a non-negligible amount of experimental data on charge-dependent effects in particle spectra and anisotropic flow exists, and much more can be obtained from existing fixed-target as well as collider experiments. These data can be used to trace the influence of the electric and magnetic fields in heavy ion collisions, which should be useful in future studies related to the chiral magnetic effect,
%~\cite{cme,cme1}, 
the electromagnetic properties of the quark-gluon plasma,
%~\cite{kharzeev2014,voloshin2014,hirono2012}, 
and others. Our own studies demonstrate the sensitivity of the EM-induced distortions of charged particle spectra and directed flow to the space-time scenario of particle production in heavy ion collisions, and allow us to trace the longitudinal evolution of the expanding matter created in the course of the collision.\\

%%%%%%%%%%%%%%%%%%%%%%%%%%%%%%
\vspace*{0.1cm}{\Large\bf Acknowledgments}\\

%The authors warmly thank Yadav Pandit and the STAR Collaboration for 
%providing the numerical values for the published STAR data.
 The authors, and especially A.R., gratefully thank the organizers of the X Workshop on Particle Correlations and Femtoscopy (WPCF 2014), for their invitation and for the excellent organization of such a fruiful and interesting Workshop.

 This work was supported by the Polish National Science Centre 
(on the basis of decision no. DEC-2011/03/B/ST2/02634).


\begin{thebibliography}{10}

%%%%%%%%%%%%%%%%%%%%%%%%%%%%
\bibitem{Rybicki-coulomb}   %2
 A. Rybicki, A. Szczurek, Phys.\ Rev.\ C {\bf 75}, 054903 (2007)
  [nucl-th/0610036].

\bibitem{kharzeev2014} 
  U.~G\"ursoy, D.~Kharzeev and K.~Rajagopal,
  %``Magnetohydrodynamics, charged currents and directed flow in heavy ion collisions,''
  Phys.\ Rev.\ C {\bf 89}, 054905 (2014)
  [arXiv:1401.3805 [hep-ph]].
  %%CITATION = ARXIV:1401.3805;%%
  %2 citations counted in INSPIRE as of 27 May 2014

\bibitem{voloshin2014} 
  V.~Voronyuk, V.~D.~Toneev, S.~A.~Voloshin and W.~Cassing,
  %``Charge-dependent directed flow in asymmetric nuclear collisions,''
  Phys.\ Rev.\ C {\bf 90}, no. 6, 064903 (2014)
  [arXiv:1410.1402 [nucl-th]].


\bibitem{hirono2012} 
  Y.~Hirono, M.~Hongo and T.~Hirano,
  %``Estimation of electric conductivity of the quark gluon plasma via asymmetric heavy-ion collisions,''
  Phys.\ Rev.\ C {\bf 90}, no. 2, 021903 (2014)
  [arXiv:1211.1114 [nucl-th]].

\bibitem{workshop}
%workshop or STAR
  L.~Adamczyk {\it et al.}  [STAR Collaboration],
  %``Beam-energy dependence of charge separation along the magnetic field in Au+Au collisions at RHIC,''
  Phys.\ Rev.\ Lett.\  {\bf 113}, 052302 (2014)
  [arXiv:1404.1433 [nucl-ex]].

\bibitem{cme}
  D.~E.~Kharzeev, L.~D.~McLerran and H.~J.~Warringa,
  %``The Effects of topological charge change in heavy ion collisions: 'Event by event P and CP violation',''
  Nucl.\ Phys.\ A {\bf 803}, 227 (2008)
  [arXiv:0711.0950 [hep-ph]].

\bibitem{cme1}
  K.~Fukushima, D.~E.~Kharzeev and H.~J.~Warringa,
  %``The Chiral Magnetic Effect,''
  Phys.\ Rev.\ D {\bf 78}, 074033 (2008)
  [arXiv:0808.3382 [hep-ph]].


\bibitem{Rybicki-epshep}
%Strong and Electromagnetic Interactions at SPS Energies, 
A.~Rybicki, PoS(EPS-HEP 2009) 031.

\bibitem{ismd2012} 
  M.~Klusek-Gawenda, E.~Kozik, A.~Rybicki, I.~Sputowska and A.~Szczurek,
  %``Strong and Electromagnetic Forces in Heavy Ion Collisions,''
  Acta Phys.\ Polon.\ Supp.\  {\bf 6}, 451 (2013)
  [arXiv:1303.6423 [nucl-ex]], and references therein.

\bibitem{wa98}   %
  H.~Schlagheck (WA98 Collaboration),
  %``Thermalization and flow in 158-GeV/A Pb + Pb collisions,''
  Nucl.\ Phys.\ A {\bf 663}, 725 (2000)
   [nucl-ex/9909005].
  %%CITATION = NUCL-EX/9909005;%%
  %3 citations counted in INSPIRE as of 20 Mar 2013

\bibitem{star2014} 
  L.~Adamczyk {\it et al.}  (STAR Collaboration),
  %``Beam-Energy Dependence of Directed Flow of Protons, Antiprotons and Pions in Au+Au Collisions,''
  Phys.\ Rev.\ Lett.\  {\bf 112}, 162301 (2014)
  [arXiv:1401.3043 [nucl-ex]].
  %%CITATION = ARXIV:1401.3043;%%
  %4 citations counted in INSPIRE as of 27 May 2014

\bibitem{Rybicki-v1} 
  A.~Rybicki and A.~Szczurek,
  %``Spectator induced electromagnetic effect on directed flow in heavy ion collisions,''
  Phys.\ Rev.\ C {\bf 87}, 054909 (2013)
  [arXiv:1303.7354 [nucl-th]],
  and references therein.
  %%CITATION = ARXIV:1303.7354;%%
  %3 citations counted in INSPIRE as of 12 Mar 2014

\bibitem{Rybicki-meson2014} 
  A.~Rybicki, A.~Szczurek and M.~Klusek-Gawenda,
  %``New results on Coulomb effects in meson production in relativistic heavy ion collisions,''
  EPJ Web Conf.\ {\bf 81}, 05024 (2014).

\bibitem{Ryblewski}
 R.~Ryblewski, private communication. 

\bibitem{na61-2014} 
  N.~Abgrall {\it et al.} (NA61/SHINE Collaboration),
  %``Measurement of negatively charged pion spectra in inelastic p+p interactions at $p_{lab}$ = 20, 31, 40, 80 and 158 GeV/c,''
  Eur.\ Phys.\ J.\ C {\bf 74}, no. 3, 2794 (2014)
  [arXiv:1310.2417 [hep-ex]], and references therein.

\end{thebibliography}
\end{document}